\documentstyle[12pt] {article}
\newcommand{\beq}{\begin{equation}}
\newcommand{\eeq}{\end{equation}}
\newcommand{\beqa}{\begin{eqnarray}}
\newcommand{\eeqa}{\end{eqnarray}}
\newcommand{\ba}{\begin{array}}
\newcommand{\ea}{\end{array}}
\begin{document}

{\begin{center}
{\large \bf Chaos suppression in \\
the SU(2) Yang--Mills--Higgs system} \\
\end{center}}

\vskip 1. truecm

\begin{center}
{\bf Luca Salasnich}\footnote{Electronic address:
salasnich@padova.infn.it}
\vskip 0.5 truecm
Dipartimento di Fisica "G. Galilei" dell'Universit\`a di Padova, \\
Via Marzolo 8, I 35131 Padova, Italy \\
\vskip 0.5 truecm
Departamento de Fisica Atomica, Molecular y Nuclear \\
Facultad de Ciencias Fisicas, Universidad "Complutense" de Madrid, \\
Ciudad Universitaria, E 28040 Madrid, Spain\\
\end{center}

\vskip 1. truecm

\begin{center}
{\bf Abstract}
\end{center}
\vskip 0.5 truecm
\par
We study the classical chaos--order transition in the spatially
homogenous SU(2) Yang--Mills--Higgs system by using a quantal analog
of Chirikov's resonance overlap criterion.
We obtain an analytical estimation of the
range of parameters for which there is chaos suppression.

\vskip 1. truecm
\begin{center}
\end{center}

\newpage

\par
In the last years there has been much interest in the chaotic behaviour
of classical field theories. In particular have been studied Yang--Mills [1,2],
Yang--Mills--Higgs [3,4,5], Abelian--Higgs [6],
and also Chern--Simons [7] systems.
\par
Usually the order--chaos transition in these systems has been studied
numerically with Lyapunov exponents [8] and sections of Poincar\`e [9].
Less attention has been paid to analytical criteria.
Some authors [4,6] have used the curvature criterion of
potential energy [10], but this
criterion guarantees only a local instability and can give incorrect
results (for a fuller discussion of this point see [11]).
\par
In this work we study analytically the suppression of classical chaos
in the spatially homogenous SU(2) Yang--Mills--Higgs (YMH) system.
We apply a quantal analog [12] of the Chirikov resonance
criterion [13] by using the semiclassical quantization
to calculate the critical value of the parameters corresponding to the
intersection of two neighboring quantal separatrices [20].
\par
Obviously, the constant field approximation implies that our SU(2) YMH
system is a toy model for classical non--linear dynamics,
with the attractive feature that the model emerges from particle physics.
\par
The lagrangian density for the SU(2) YMH system is given by [14]:
\beq
L=-{1\over 4}F_{\mu \nu}^{a}F^{\mu \nu a}
+{1\over 2}(D_{\mu}\phi )^+(D^{\mu}\phi )
-V(\phi ) \; ,
\eeq
where:
\beq
F_{\mu \nu}^{a}=\partial_{\mu}A_{\nu}^{a}-\partial_{\nu}A_{\mu}^{a}+
g\epsilon^{abc}A_{\mu}^{b}A_{\nu}^{c} \; ,
\eeq
\beq
(D_{\mu}\phi )=\partial_{\mu}\phi - i g A_{\mu}^b T^b\phi
\; ,
\eeq
with $T^b=\sigma^b/2$, $b=1,2,3$, generators of the SU(2) algebra,
and where the potential of the scalar field (the Higgs field) is:
\beq
V(\phi )=\mu^2 |\phi|^2 + \lambda |\phi|^4 \; .
\eeq
We work in the (2+1)--dimensional Minkowski space ($\mu =0,1,2$) and
choose spatially homogenous Yang--Mills and the Higgs fields:
\beq
\partial_i A^a_{\mu} = \partial_i \phi = 0, \;\; i=1,2
\eeq
i.e. we consider the system in the region where space fluctuations of
fields are negligible compared to their time fluctuations.
\par
In the gauge $A^a_0=0$ and using the real triplet representation for the
Higgs field we obtain:
$$
L={1\over 2}({\dot {\vec A_1}}^2+{\dot {\vec A_2}}^2)+
{\dot {\vec \phi}}^2
-g^2 [{1\over 2}{\vec A_1}^2 {\vec A_2}^2
-{1\over 2} ({\vec A_1} \cdot {\vec A_2})^2+
$$
\beq
+({\vec A_1}^2+{\vec A_2}^2){\vec \phi}^2
-({\vec A_1} \cdot {\vec \phi})^2 -({\vec A_2} \cdot {\vec \phi})^2]
-V({\vec \phi}),
\eeq
where ${\vec \phi}=(\phi^1,\phi^2,\phi^3)$,
${\vec A_1}=(A_1^1,A_1^2,A_1^3)$, and
${\vec A_2}=(A_2^1,A_2^2,A_2^3)$.
\par
When $\mu^2 >0$ the potential $V$ has a minimum in $|{\vec \phi}|=0$,
but for $\mu^2 <0$ the minimum is:
$$
|{\vec \phi_0}|=({-\mu^2\over 4\lambda })^{1\over 2}=v
$$
which is the non zero Higgs vacuum. This vacuum is degenerate
and after spontaneous symmetry breaking the physical vacuum can be
chosen ${\vec \phi_0} =(0,0,v)$.
If $A_1^1=q_1$, $A_2^2=q_2$, and the other components of the
Yang--Mills fields are zero, in the Higgs vacuum the hamiltonian of the
system is:
\beq
H={1\over 2}(p_1^2+p_2^2)
+g^2v^2(q_1^2+q_2^2)+{1\over 2}g^2 q_1^2 q_2^2 \; ,
\eeq
where $p_1={\dot q_1}$ and $p_2={\dot q_2}$. Obviously $w^2=2 g^2v^2$ is the
mass term of the Yang--Mills fields.
\par
Classical chaos was demonstrated in a pure Yang--Mills system [1], i.e. in a
zero Higgs vacuum. Here we analyze the effect of a non zero
Higgs vacuum [3].
\par
We introduce the action--angle variables
by the canonical transformation:
\beq
q_i=({2I_i\over \omega})^{1\over 2}\cos{\theta_i} \;, \;\;\;\;
p_i=({2I_i \omega})^{1\over 2}\sin{\theta_i} \; , \;\; i=1,2.
\eeq
The hamiltonian becomes (see also [3]):
\beq
H=(I_1+I_2)\omega +{1\over v^2} I_1I_2
       \cos^2{\theta_1}\cos^2{\theta_2}.
\eeq
By the new canonical transformation in slow and fast variables:
$$
A_1=I_1+I_2 \; , \;\;\; A_2=I_1-I_2 \; ,
$$
\beq
\theta_1=\chi_1+\chi_2 \; , \;\;\; \theta_2=\chi_1-\chi_2 \; ,
\eeq
$H$ can be written:
\beq
H=A_1\omega + {1\over 4 v^2}(A_1^2-A_2^2)
       \cos^2{(\chi_1+\chi_2)}\cos^2{(\chi_1-\chi_2)}.
\eeq
We now eliminate the dependence on the angles to order $1/v^4$ by
resonant canonical perturbation theory [15]. First we average
on the fast variable $\chi_1$. This yields:
\beq
{1\over 2\pi}\int_{0}^{2\pi}d\chi_{1}
\cos^{2}{(\chi_{1}+\chi_{2})}\cos^{2}{(\chi_{1}-\chi_{2})} ={1\over
8}(2+\cos{4\chi_2}),
\eeq
and:
\beq
{\bar H}_{cl}=A_1\omega +{1 \over 32 v^2}(A_{1}^{2}-A_{2}^{2})
(2+\cos{4\chi_2}).
\eeq
The dependence on $\chi_2$ is now eliminated by a second canonical
transformation. The Hamilton--Jacobi equation for the perturbation part
is indeed:
\beqa [A_{1}^{2}-({\partial S\over \partial
\chi_{2}})^{2}] (2+\cos{4\chi_{2}})=K,
\\
{\partial S\over \partial \chi_{2}}=\pm
\sqrt{ A_{1}^{2}(2+\cos{4\chi_{2}})-K\over 2+\cos{4\chi_{2}} }.
\eeqa
and thus the Hamiltonian becomes:
\beq
{\bar H}=B_{1}+{1 \over 32 v^2}K(B_{1},B_{2}),
\eeq
where:
\beq
B_{1}=A_{1}, \;\;\;\;
B_{2}={1\over 2\pi}\oint d\chi_2 {\partial S\over \partial \chi_2}.
\eeq
It appears from the structure of this equation that the motion of our
system is similar that of a simple pendulum:
for $0<K<B_1^2$ rotational motion, for $K=B_{1}^{2}$
separatrix, and for $B_{1}^{2}<K<3B_{1}^{2}$ librational motion.
On the separatrix we have $B_{1}^{2}(2+\cos{4\chi_{2}})=K$, and:
\beq
B_2=\pm {2\over \pi}\int_{a}^{b}dx
\sqrt{ B_{1}^{2}(2+\cos{4x})-K\over 2+\cos{4x} },
\eeq
where $a=-{\pi \over 4}$, $b={\pi\over 4}$ for rotational motion, and
$a=\phi_{-}(K,B_{1})$, $b=\phi_{+}(K,B_{1})$ for librational motion, with:
\beq
\phi_{\pm}(K,B_{1})=\pm {1\over 4}\arccos ({K\over B_1^2}-2).
\eeq
The appearance of a separatrix (which is not immediately obvious in the
$(p,q)$ coordinates) accounts as is well known for the
stochastic layers originating near it [16]. This corresponds to local irregular
behaviour of the quantum spectrum; one of its manifestations is
the local shrinking of the level spacing and the tendency
to avoided crossing [16,17].
\par
The approximate hamiltonian (16) depends only on the
actions so that a semiclassical quantization formula
can be obtained by the Bohr--Sommerfeld
quantization rules [15,18]. Set $B_1=m_1\hbar$ and $B_2=m_2\hbar$, then,
up to terms of order $\hbar$, the quantum spectrum is:
\beq
E_{m_{1},m_{2}}=m_{1}\hbar\omega +{1 \over 32 v^2}K(m_{1}\hbar , m_{2}\hbar ),
\eeq
where $K$ is implicitly defined by the relation:
\beq
m_{2}\hbar =\pm {2\over \pi}\int_{a}^{b}dx
\sqrt{ (m_{1}\hbar )^{2}(2+\cos{4x})-K\over 2+\cos{4x} },
\eeq
with $a=-{\pi\over 4}$, $b={\pi\over 4}$ for $0<K<(m_{1}\hbar )^2$, and
$a=\phi_{-}(K,B_{1})$, $b=\phi_{+}(K,B_{1})$ for
$(m_{1}\hbar )^{2}<K<3(m_{1}\hbar )^{2}$.
\par
On the separatrix, where $K=(m_1\hbar )^2$, $m_2 =\pm \alpha m_1$,
with:
\beq
\alpha ={2\over \pi}\int_{-{\pi\over 4}}^{\pi\over 4}
dx \sqrt{1+\cos{4x}\over 2+\cos{4x}}.
\eeq
\par
It is immediate to see that for $m_1$ fixed the function $K$, and
hence the semiclassical energy $E_{m_1,m_2}$, is a decreasing function of
the secondary quantum number $m_2$, and we have a quantum multiplet [19].
\par
We can calculate the value of the coupling
constant $1/v^2$ corresponding to the intersection of the
separatrices of two neighboring quantum multiplets:
\beq
(m_1+1)\hbar\omega +{1 \over 32 v^2}K[(m_1+1)\hbar ,\alpha (m_1+1)\hbar ]=
m_1\hbar\omega +{1 \over 32 v^2}K(m_1\hbar ,\alpha m_1\hbar ),
\eeq
and so:
\beq
{1\over v^2}={ -32 \hbar\omega \over
K[(m_1+1)\hbar ,\alpha (m_1+1)\hbar ]-K(m_1\hbar ,\alpha m_1\hbar ) }.
\eeq
In this way we have, in some sense, the quantal
counterpart [12] of the method of overlapping resonances developed by
Chirikov [13]. The denominator can be evaluated by the Taylor expansion and
finally:
\beq
{1\over v^2}= \left[ { -8\omega \over
{\partial K\over \partial B_{1}}-
\alpha{\partial K \over \partial B_{2}} }
\right]_{B_1=m_1\hbar ,B_2=\alpha m_2\hbar}.
\eeq
K is implicitly defined by the relation:
\beq
F[B_1,B_2,K(B_1,B_2)]=B_2-{\pi\over 2}
\int_{-{\pi\over 4}}^{\pi\over 4}dx
\sqrt{ B_1^{2}(2+\cos{4x})-K\over 2+\cos{4x} }=0,
\eeq
or:
\beq
F(B_{1},B_{2},K)=B_{2}-\Phi (B_{1},K)=0.
\eeq
As a function of $\Phi$, $1/v^2$ can be written:
\beq
{1\over v^2}= \lim_{K\to B_1^2}
\left[ { 8\omega {\partial \Phi\over \partial K}
\over
\alpha -{\partial \Phi\over \partial B_{1}}}
\right]_{B_1=m_1\hbar},
\eeq
where:
$$
{\partial \Phi\over \partial K}=-{1\over \pi}
\int_{-{\pi\over 4}}^{\pi\over 4} dx
{ 1\over \sqrt{(2+\cos{4x})[B_1^2(2+\cos{4x})-K]} }
$$
\beq
{\partial \Phi\over \partial B_1}={2\over \pi}
\int_{-{\pi\over 4}}^{\pi\over 4}dx
\sqrt{ {B_1^2(2+\cos{4x})\over B_1^2(2+\cos{4x})-K} }.
\eeq
A similar procedure has been used for a more schematic model in [20].
The result is:
\beq
{1\over v^2}={16 \omega \over m_1\hbar} \; ,
\eeq
where $m_1\hbar \simeq E$ (the energy of the system) and $\omega = (2 v^2
g^2 )^{1\over 2}$.
Therefore the chaos--order transition depends by the parameter
$\lambda = v^3 g/E$:
if $0 < \lambda < \sqrt{2}/32$ a relevant region of the
phase--space is chaotic, but if $\lambda > \sqrt{2}/32$ the system
becomes regular. This result shows that
the value of the Higgs field in the vacuum $v$ plays an important role:
for large values makes the system regular, in agreement with previous numerical
calculations [3]. Also the Yang--Mills coupling constant $g$ has the
same role. Instead if $v$ and $g$ are fixed there
is an order--chaos transition increasing the energy $E$.
\par
In conclusion, we have shown for the spatially homogenous SU(2) YMH system
that the quantum resonance criterion,
which describes the onset of widespread chaos associated to
semiclassical crossing between separatices of different quantum multiplets,
gives an analytical estimation of the classical chaos--order transition
as a function of the higgs vacuum, the Yang--Mills coupling constant
and the energy of the system.
\par
We observe that a classical chaos--order transition has been found
numerically for the monopole solution [4] and the sphaleron solution [5]
of the SU(2) YMH theory. In the future, it will be of great importance
to find analytical estimations of the onset of chaos also for these
more realistic solutions.

\vskip 0.5 truecm

\begin{center}
*****
\end{center}
\par
The author is greatly indebted to Profs. G. Benettin, S. Graffi,
V. R. Manfredi and Dr. A. Vicini for many enlightening discussions.
The author acknowledges Prof. J. M. G. Gomez for his kindly
hospitality at the Department of Atomic, Molecular and Nuclear Physics
of "Complutense" University, and the "Ing. Aldo Gini" Foundation
for a partial support.

\newpage
\begin{center}
{\bf References}
\end{center}
\vskip 0.5 truecm

[1] G. K. Savvidy: Phys. Lett. B {\bf 130}, 303 (1983)

[2] G. K. Savvidy: Phys. Lett. B {\bf 159}, 325 (1985)

[3] G. K. Savvidy: Nucl. Phys. B {\bf 246}, 302 (1984)

[4] T. Kawabe, S. Ohta: Phys. Rev. D {\bf 44}, 1274 (1991)

[5] T. Kawabe, S. Ohta: Phys. Lett. B {\bf 334}, 127 (1994)

[6] T. Kawabe: Phys. Lett. B {\bf 343}, 254 (1995)

[7] M. S. Sriram, C. Mukku, S. Lakshmibala, B. A. Bambah: Phys. Rev. D
{\bf 49}, 4246 (1994)

[8] A. J. Lichtenberg, M. A. Lieberman: {\it Regular and Stochastic
Motion} (Springer--Verlag, 1983)

[9] H. Poincar\`e: {\it New Methods of Celestial Mechanics}, vol. 3, ch.
27 (Transl. NASA Washington DC 1967); M. Henon: Physica D {\bf 5}, 412 (1982)

[10] M. Toda: Phys. Lett. A {\bf 48}, 335 (1974)

[11] G. Benettin, R. Brambilla, L. Galgani: Physica A {\bf 87}, 381 (1977)

[12] S. Graffi, T. Paul, H. J. Silverstone: Phys. Rev. Lett. {\bf 59}, 255
(1987)

[13] B. V. Chirikov: Phys. Rep. {\bf 52}, 263 (1979)

[14] C. Itzykson, J. B. Zuber: {\it Quantum Field Theory} (McGraw--Hill, 1985)

[15] M. Born: {\it Mechanics of the Atom} (Bell, 1960);
J. Bartels and S. J. Chang: Phys. Rev. A {\bf 41}, 598 (1990)

[16] M. C. Gutzwiller: {\it Chaos in Classical and Quantum Mechanics}
(Springer--Verlag, 1991)

[17] D. Delande: {\it Chaos and Quantum Physics}, in Les
Houches Summer School 1989, Ed. M. J. Giannoni, A. Voros and J. Zinn--Justin
(Elsevier Science Publishing, 1989)

[18] V. P. Maslov and M. V. Fedoriuk: {\it Semi--Classical Approximation in
Quantum Mechanics} (Reidel Publishing Company, 1981);
W. P. Reinhardt: in {\it The Mathematical Analysis of Physical Systems},
Ed. R. E. Mickens and R. Van Nostrand (1984)

[19] S. Graffi, V. R. Manfredi, L. Salasnich:
Nuovo Cim. B {\bf 109}, 1147 (1994)

[20] S. Graffi, V. R. Manfredi, L. Salasnich:
Mod. Phys. Lett. B {\bf 7}, 747 (1995)

\end{document}